\documentclass{ws-procs975x65}

\usepackage{colordvi}
\usepackage{color}
\usepackage{amsfonts, amsmath, amssymb,calligra}

\begin{document}
	
	\title{Local rotational symmetry Gowdy model in Loop Quantum Gravity}

\author{Javier Olmedo$^1$, Daniel Mart\'{\i}n de Blas$^2$, Tomasz Paw{\l}owski$^{3,4}$}
\address{1. Department of Physics and Astronomy, Louisiana State University,
	Baton Rouge, LA 70803-4001\\
	2. Departamento de Ciencias F\'isicas, Facultad de Ciencias Exactas, \\
	Universidad Andr\'es Bello, Av. Rep\'ublica 220, Santiago 8370134, Chile\\
        3. Center for Theoretical Physics, Polish Academy of Sciences, \\
        Al. Lotnik\'ow 32/46, 02-668 Warsaw, Poland\\
        4. Instytut Fizyki Teoretycznej, Uniwersytet Warszawski, \\
        Pasteura 5, 02-093 Warszawa, Poland, EU        
      }

\begin{abstract}
  We provide a complete quantization for the Gowdy model with local rotational symmetry in vacuum. We start with a redefinition of the classical constraint algebra such that the Hamiltonian constraint has a vanishing Poisson bracket with itself. We apply a canonical quantization within loop quantum gravity and an improved dynamics scheme. We construct the exact solutions to the constraints and the physical Hilbert space, together with the physical observables. The quantization provides a physical picture without singularities. Besides, a genuine discretization of the spatial geometry emerges by means of a 
  new quantum observable without classical analogue.
\end{abstract}

The study of non-trivial solutions of Einstein's theory admitting cosmological spacetimes has been very fruitful since they allowed to understand many physical phenomena. Among them, Gowdy spacetimes \cite{gowdy1} provide a suitable arena since, even in vacuum, they admit cosmological solutions like Bianchi cosmologies with non-perturbative gravitational waves. Different aspects of  the quantization of these spacetimes have been studied in the past years, even the problem of the cosmological singularity \cite{hybrid}. In that reference, the authors considered the flat linearly polarized Gowdy model (with three torus topology). They proposed the so-called hybrid quantization that combines a loop quantization for the homogeneous degrees of freedom with a standard Fock representation for the gravitational waves. After performing a partial gauge fixing, one is left with a global Hamiltonian and diffeomorphism constraints. The solutions of the system can be determined and the physical Hilbert space constructed out of them. The classical singularity is then suitably eliminated. On the other hand, a complementary study in terms of Ashtekar--Barbero variables \cite{baren} shows  the difficulties for performing a full canonical quantization using conventional loop quantum gravity techniques.  

Here we will not consider the full vacuum polarized Gowdy three torus model, but its locally rotational symmetry (LRS) version ---it is a preliminary step that can seed some light on the quantization of the full model---. In this case, one imposes to the two homogeneous directions to be identical. The model is simpler, since the classical inhomogeneous model possesses only one (global) physical degree of freedom, while it is still physically relevant since its coupling to a massless scalar field \cite{LRS-gowdy} admits homogeneous and isotropic solutions. Our purpose is to proceed with a quantization without gauge-fixing the system. 
We will follow the strategy already adopted for the study of spherically symmetric spacetimes \cite{BH-2}. There, one redefines the scalar constraint in such a way that the new one has an Abelian Poisson algebra with itself, and the usual one with the diffeomorphism constraint. 
In addition, we adhere to an improved dynamics scheme. We provide the solutions to the constraints and endow them with Hilbert space structure. It is remarkable that the resulting spacetimes are free of singularities. Besides, the area of the Killing orbits is quantized due to the discreteness of the spectrum of a kinematical operator, that becomes a physical observable, without classical analog. 

Let us start by adopting a description in terms of Ashtekar--Barblero variables \cite{baren}. Besides, and for the sake of simplicity, we will assume that the Immirzi parameter is equal to the unit. The variables $E^x$, ${K}_x$ and $\cal E$, $\cal A$, (densitized triads and connections) are the coordinates of our phase space. They take values 
on the circle, i.e. $\theta\in [0,2\pi )$. The spatial metric components are $g_{\theta\theta}=(E^{x})^2\mathcal{E}^{-1}$, $g_{xx}=g_{yy}=\mathcal{E}$, where the remaining spatial coordinates are $\{x,y\}\in [0,2\pi )$. The triad $\mathcal{E}$ can be identified with the area of the spatial Killing vectors. The remaining components of the metric can be written in terms of the lapse $N$ and the shift $N^{\theta}$. 

The total Hamiltonian $H_T$ of this model is a linear combination of constraints without a true Lie algebra. We proceed to redefine the scalar constraint by introducing the new lapse $\tilde N=NE^x$ and shift $\tilde N_\theta=N_\theta+2NK_x\sqrt{{\cal E}}(\partial_\theta{\cal E})^{-1}$, such that
\begin{equation}
H_T =\frac{1}{\kappa}\int d\theta\;
\frac{\tilde N}{(\partial_\theta{\cal E})}\partial_\theta\left[-2\sqrt{{\cal E}}K_x^2+\frac{\sqrt{\cal E}(\partial_\theta{\cal E})^2}{2(E^x)^{2}}\right]+2\tilde N_\theta
\left[ (\partial_{\theta}K_x)E^x - (\partial_{\theta}\cal{E})\cal{A} \right].
\end{equation}
It is worth emphasizing that the scalar constraint corresponding to the new lapse $\tilde N$  has the usual Poisson algebra with the diffeomorphism constraint and commutes, under Poisson brackets, with itself.

Now we will proceed with the loop quantization of the model. The kinematical Hilbert space possesses a basis of one-dimensional spin-networks $|\vec{k},\vec{\mu}\rangle$. They are cylindrical functions of the connection consisting of holonomies of the connection along disjoint edges $e_j$, with valences $k_j\in \mathbb{N}\cup\{0\}$, joined at the end points by vertices $v_j$ with valences $\mu_j\in \mathbb{R}$.
Our purpose is to implement an improved dynamics scheme.  So it is more appropriate to consider a state labeling $\mu_{j} \rightarrow \nu_{j}=\sqrt{k_{j}}\mu_{j}/\lambda$, with $\lambda$ a dimensionless real parameter proportional to the minimum allowed area. The inner product in this basis is
$\langle\vec{k},\vec{\nu}|\vec{k}',\vec{\nu}'\rangle=\delta_{\vec{k}\vec{k}'}\delta_{\vec{\nu}\vec{\nu}'},$
keeping in mind that different graphs are mutually orthogonal. The basic operators act as ${\hat{\cal E}(\theta) } |\vec{k},\vec{\nu}\rangle
=\ell_{\rm Pl}^2 k_j |\vec{k},\vec{\nu}\rangle$ and $\hat{V}(\theta) |\vec{k},\vec{\nu}\rangle
=\lambda \ell_{\rm Pl}^3 \sum_{v_j\in g} \delta(\theta-\theta(v_j))\nu_j 
|\vec{k},\vec{\nu}\rangle$, where $\hat{V}=\hat{\sqrt{\cal E}}\hat E^x$ is the volume operator and $\ell_{\rm Pl}^2=G\hbar.$ 

We now proceed to polymerize the corresponding contribution in the Hamiltonian constraint within an improved dynamics scheme. It introduces a minimum ``length'' for the point holonomies such that $\rho^{2}_{j}k_{j}=\lambda^{2}$. One of the simplest choices is $K_x\to\sin\left(\rho_{j}
  K_x\right)/\rho_{j}$. On a given graph, the scalar constraint operator is defined as
\begin{equation}\label{eq:quant-scalar-constrh0}
\hat{H}(N)=\sum_{j=1}^n N_j\hat P\frac{1}{\ell_{\rm Pl}\Delta k_j}\left[ (k_{j-1})^{3/2}\hat{ h}_{j-1}-(k_{j})^{3/2}\hat{h}_{j}\right]\hat P,
\end{equation}
where $\hat P|\vec{k},\vec{\nu}\rangle=\prod_{v_j}\left[{\rm sgn}(k_j){\rm sgn}(\nu_j)\right]^2|\vec{k},\vec{\nu}\rangle$ has been introduced for convenience in order to decouple vertices with $k_j=0$ and/or $\nu_j=0$. Besides, we have defined the eigenvalue of $\widehat{\partial_\theta \cal E}$ as $\ell_{\rm Pl}^{2}\Delta k_{j}$, with $\Delta k_j=(k_{j}-k_{j-1})$. Regarding the operator
\begin{equation}
\hat h_j=\widehat{\left[\frac{1}{V}\right]}_j^{1/2}\left(2  
\hat\Omega^2_j
-\frac{1}{2}\widehat{\left[\frac{1}{V}\right]}_j \ell_{\rm Pl}^{2}(\ell_{\rm Pl}^2\Delta k_j)^2\right)\widehat{\left[\frac{1}{V}\right]}_j^{1/2},
\end{equation}
we have adopted a regularization of the inverse triad operators \emph{\`a la} Thiemann such that $\widehat{\left[\frac1{V}\right]}_j^{1/2} |\vec{k},\vec{\nu}\rangle = \frac{1}{\ell_{\rm Pl}^{3/2}\lambda^{1/2}}\left||\nu_j+1|^{1/2}-|\nu_j-1|^{1/2}\right||\vec{k},\vec{\nu}\rangle$, 
and
\begin{equation}
\hat{\Omega}_j= \frac{1}{{4i\lambda}}|\hat{V}|^{1/4}\big[\widehat{{\rm sgn}(V)}\big(\hat{\mathcal{N}}^x_{2\bar{\rho}_{j}}-\hat{\mathcal{N}}^x_{-2\bar{\rho}_{j}}\big)+\big(\hat{\mathcal{N}}^x_{2\bar{\rho}_{j}}-\hat{\mathcal{N}}^x_{-2\bar{\rho}_{j}}\big)\widehat{{\rm sgn}(V)}\big]|\hat{V}|^{1/4}\Big|_{\theta=\theta(v_j)},
\end{equation}
with $(\hat{\mathcal{N}}^x_{\pm 2\bar{\rho}_{j}})_{\theta=\theta_j} |k_j,\nu_j\rangle=|k_j,\nu_j\pm2\rangle$. 

Let us comment that a similar operator has been already studied in spherically symmetric spacetimes \cite{BH-2}. It is important to notice that: i) it commutes with itself; ii) the states with any $k_j=0$ or $\nu_j=0$, or both, are trivially annihilated by the constraints (the quantum analogues to the classical singularity can be decoupled from the theory since they are irrelevant in the dynamics); iii) states with $\Delta k_j=0$ yields an ill defined quantum constraint, so they do not belong to its domain of  definition; iv) the solutions  admit a natural factorization on each vertex; v) at each vertex $v_j$ the scalar constraint acts as a difference operator in the label $\nu_j$ such that it decouples states with $\nu_j>0$ with respect to those with $\nu_j<0$, and allows to restrict the study to superselections sectors $\mathcal{L}_{\epsilon_{j}}=\{\nu_{j} | \nu_{j}=\epsilon_{j}+4m,\ m\in \mathbb{N},\ \epsilon_{j}\in(0,4]\}$.

If the solutions  to the scalar constraint are assumed to be of the form of $(\Psi|=\sum_{\vec k}\sum_{\vec \nu}\langle \vec{k},\vec{\nu}|\psi(\vec k,\vec{\nu})$, due to iv), it is straightforward to see by direct inspection that the solutions at each vertex must be annihilated by the operator $[\hat{h_j}^\dagger-h_0(k_{j})^{-3/2}]$
with $h_0$ a global constant (physically meaningful for the cosmological sector if it takes positive values). In order to solve the scalar constraint on that sector it is enough to consider the positive spectrum of $\hat h_j^\dagger$, that is continuous and nondegenerated. The solutions can then be obtained by group averaging techniques. The resulting states $(\tilde \Psi_g|_{h_0}$ depend on $h_0$. This parameter is in fact an observable of the model since it parametrizes different physical solutions in the classical theory (one can see that there $h_0$ is a Dirac observable). In consequence, it is natural to consider instead superpositions of it rather than particular solutions for given values of $h_0$. The inner product is then 
\begin{equation}\label{eq:phys-inner-prod}
(\tilde \Psi|\tilde\Phi\rangle=\int_{0}^{h_{*}}dh\sum_{\vec k}(\tilde\psi(\vec k,\vec{\omega}_h,h))^*\tilde\phi(\vec k,\vec{\omega}_h,h):=\int_{0}^{h_{*}}dh\sum_{\vec k}(\tilde\psi(\vec k,h))^*\tilde\phi(\vec k,h), 
\end{equation}
where $h_{*}=\min[\vec k]^{3/2}/\lambda^2$. Finally, we require them to be invariant under spatial diffeomorphisms (the symmetries generated by the diffeomorphism constraint). We then follow the usual strategy in loop quantum gravity that averages the graphs with respect to the group of finite diffeormophisms. 
The physical states of the theory correspond to those $(\tilde \Psi|$ with vertices in all possible positions along the $\theta$-direction with their order preserved. Besides, the number of vertices as well as the quantum numbers $h$ and $\vec{k}$ are preserved under both the scalar constraint and the spatial diffeomorphisms.

Regarding the observables of the model, we have identified two basic quantities. The first one is a global degree of freedom codified by $\hat h$. The second observable corresponds to the sequence of integers $\vec{k}$. As it was noticed in spherically symmetric vacuum spacetimes, this observable can be promoted to a parametrized observable defined as $  \hat{O}(z)\vert \vec{k},h\rangle_{\rm phys}= \ell_{\rm Planck}^2 k_{{\rm Int}(nz)} \vert \vec{k},h\rangle_{\rm phys}$ where $\vert  \vec{k},h\rangle_{\rm phys}$ is a basis of eigenstates of the observables, $n$ is the number of vertices, $z\in [0,1]$ and ${\rm Int}(n z)$ is the integer part of $n z$. This observable has no classical analog. It codifies the sequence of areas of the Killing orbits. For instance, $g_{xx}(t,\theta)=g_{yy}(t,\theta)={\cal E}(t,\theta)$
are not Dirac observables since they are not invariant under a spatial diffeomorphism but 
they  can be defined as parametrized observables by means of $\hat{\cal E}(t,\theta) \vert \vec{k},h\rangle_{\rm phys} = \hat{O}(z(t,\theta))\vert \vec{k},h\rangle_{\rm phys} $ (with $z(t,\theta)$ as gauge function).

Let us finish the analysis by discussing the semiclassical regime of this description. It seems natural to think that smooth geometries will be provided by graphs with a sufficiently high number of vertices and small jumps of the expectation value of the components of the metric between consecutive vertices. This picture might be provided for states highly peaked around given values of the observables $\hat h$ and $\hat{O}(z)$. Among the semiclassical states, there is a relevant subclass corresponding to those ones providing (at least approximated) homogeneous cosmologies. In the classical theory, homogeneous geometries are characterized by a homogeneous metric $g_{\mu\nu}(t)$. If we consider suitable superpositions of states for a sufficiently large number of vertices together with suitable homogeneous gauge parameters, the previous requirement can be fulfilled in a very good approximation. Though, there could be different notions of homogeneity at the quantum level. All these issues together with an extension of the model for $h<0$ and a comparison with previous proposals will be a matter for future research.

We wish to thank Rodolfo Gambini, Guillermo A. Mena Marug\'an and Jorge Pullin for comments.
This work was supported in part by the Polish Narodowe Centrum Nauki (NCN) grant 2012/05/E/ST2/03308, the Chilean grant CONICYT/FONDECYT/REGULAR/1140335 and the grants MICINN/MINECO FIS2011-30145-C03-02 and FIS2014-54800-C2-2-P from Spain. D. M-dB is supported by the project CONICYT/FONDECYT/POSTDOCTORADO/3140409 from Chile. J. O. acknowledges Pedeciba and the grant NSF-PHY-1305000.

\end{document}